\documentclass[runningheads]{llncs}

\usepackage{fancyvrb} % sophisticated verbatim text
	\fvset{tabsize=4}
\usepackage{amsmath} % standard maths package
\usepackage{amssymb} % extra maths symbols, like \widehat
\usepackage{centernot} % horizontally-centered negation overlaid on symbols

\usepackage[capitalize]{cleveref}
\crefformat{equation}{(#2#1#3)}

\usepackage{enumitem} % For customizing enumeration labels

\usepackage{bussproofs} % sequent calculus
\usepackage{xcolor} % create red-coloured text for notes
	\newenvironment{notes}{\color{red}\noindent}{}
\usepackage{listings} % code listings that escape to math mode on '$'
\usepackage{algorithm}
\usepackage[noend]{algpseudocode}
\makeatletter
	\def\BState{\State\hskip-\ALG@thistlm}
	\algnewcommand{\LineComment}[1]{\Statex \hskip\ALG@thistlm #1}
	\algnewcommand{\IndentLineComment}[1]{\Statex \hskip\ALG@tlm #1}
\makeatother
\usepackage{float} % flexible floating containers

\renewcommand{\epsilon}{\varepsilon}

\newcommand{\func}[1]{\mathit{#1}}
\newcommand{\sdef}{\:\:\widehat{=}\:\:}
\newcommand{\vv}{\textbf{v}}
\newcommand{\uu}{\textbf{u}}
\newcommand{\ww}{\textbf{w}}
\newcommand{\xx}{\textbf{x}}
\newcommand{\kw}[1]{\textbf{#1}}
\newcommand{\FUNCTION}[3]{\kw{fun}\ #1(#2):#3}
\newcommand{\GHOST}[3]{\kw{ghost fun}\ #1(#2):#3}
\newcommand{\AXIOM}[2]{\kw{lemma}\ \{\kw{axiom}\}\ #1(#2)}
\newcommand{\Spec}[3]{&#2\sdef#3&\\&\text{where }#1}
\newcommand{\GhostSpec}[3]{&#2&\\&\quad\kw{ensures}\ #3\\&\text{where }#1}
\newcommand{\TotalSpec}[2]{&#1\sdef#2&}
\newcommand{\TotalGhostSpec}[2]{&#1&\\&\quad\kw{ensures}\ #2}
\newcommand{\Assert}{\kw{assert}}

\newcommand{\NeuralNetwork}{\kw{NeuralNet}}
\newcommand{\Matrix}{\kw{Matrix}}
\newcommand{\Vector}{\kw{Vector}}
\newcommand{\Seq}[1]{[#1]}
\newcommand{\True}{\kw{True}}
\newcommand{\False}{\kw{False}}
\newcommand{\Real}{\kw{real}} % consistent macro capitalisation for types

\newcommand{\Nat}{\kw{nat}}
\newcommand{\Bool}{\kw{bool}}
\newcommand{\Funname}[1]{\mathit{#1}}
\newcommand{\LTwo}{\Funname{L2}} % \L is already defined
\newcommand{\Sqrt}{\Funname{Sqrt}}
\newcommand{\Sum}{\Funname{Sum}}
\newcommand{\Apply}{\Funname{Apply}}
\newcommand{\Square}{\Funname{Square}}
\newcommand{\Minus}{\Funname{Minus}}
\newcommand{\OpNorm}{\Funname{OpNorm}}
\newcommand{\Truncate}{\Funname{Truncate}}
\newcommand{\TruncateWithError}{\Funname{TruncateWithError}}
\newcommand{\Expand}{\Funname{Expand}}
\newcommand{\FrobeniusNorm}{\Funname{FrobeniusNorm}}
\newcommand{\ApplyNeuralNet}{\Funname{ApplyNN}}
\newcommand{\ApplyNeuralNetBody}{\Funname{ApplyNNBody}}
\newcommand{\ApplyLayer}{\Funname{ApplyLayer}}
\newcommand{\MV}{\Funname{MVProduct}}
\newcommand{\MatrixDiv}{\Funname{MatrixDiv}}
\newcommand{\DotProduct}{\Funname{DotProduct}}
\newcommand{\ApplyRelu}{\Funname{ApplyRelu}}
\newcommand{\Relu}{\Funname{Relu}}
\newcommand{\Robust}{\Funname{Robust}}
\newcommand{\ArgMax}{\Funname{ArgMax}}
\newcommand{\Rows}{\Funname{Rows}}
\newcommand{\Cols}{\Funname{Cols}}
\newcommand{\CompatibleInput}{\Funname{IsInput}}
\newcommand{\IsMarginLipschitzBound}{\Funname{IsMarginLipBound}}
\newcommand{\LTwoUpperBound}{\Funname{L2UpperBound}}
\newcommand{\CompatibleOutput}{\Funname{IsOutput}}
\newcommand{\GetFirstColumn}{\Funname{GetFirstColumn}}
\newcommand{\RemoveFirstColumn}{\Funname{RemoveFirstColumn}}
\newcommand{\Distance}{\Funname{Distance}}
\newcommand{\Transpose}{\Funname{Transpose}}
\newcommand{\SumMatrixElements}{\Funname{SumMatrixElements}}
\newcommand{\SquareMatrixElements}{\Funname{SquareMatrixElements}}
\newcommand{\MatrixMatrixProduct}{\Funname{MMProduct}}
\newcommand{\AssumptionOne}{\Funname{Assumption1}}
\newcommand{\AssumptionTwo}{\Funname{Assumption2}}
\newcommand{\MMGetRow}{\Funname{MMGetRow}}
\newcommand{\GramIteration}{\Funname{GramIteration}}
\newcommand{\FrobeniusNormUpperBound}{\Funname{FrobeniusNormUpperBound}}

\newcommand{\SqrtUpperBound}{\Funname{SqrtUpperBound}}
\newcommand{\Certify}{\Funname{Certify}}
\newcommand{\AreLipBounds}{\Funname{AreLipBounds}}
\newcommand{\GenerateLipschitzBound}{\Funname{GenLipschitzBound}}
\newcommand{\Abs}{\Funname{Abs}}

\newcommand{\ITE}[3]{\kw{if}\ #1\ \kw{then}\ #2\kw{else}\ #3}

\newif\ifcomments
\commentsfalse

\newif\ifExtended
\Extendedtrue

\usepackage{tikz}
\usetikzlibrary{shapes.geometric, arrows}

\begin{document}

\title{A Formally Verified Robustness Certifier for Neural Networks\ifExtended\\(Extended Version)\fi}
% extended version title is too long so we need to use this macro
\titlerunning{A Formally Verified Robustness Certifier for Neural Networks}

\author{James Tobler\inst{2}\thanks{This work was conducted while the author was employed at University of Melbourne}\and
Hira Taqdees Syeda\inst{1} \and
Toby~Murray\inst{1}}
\authorrunning{J.~Tobler et al.}

\institute{University of Melbourne, Australia
\and
University of Queensland, Australia}

% \author{\IEEEauthorblockN{James Tobler}
% \IEEEauthorblockA{\textit{School of Computing and Information Systems} \\
% \textit{University of Melbourne}\\
% Melbourne, Australia \\
% james.tobler@unimelb.edu.au}
% \and
% \IEEEauthorblockN{Toby Murray}
% \IEEEauthorblockA{\textit{School of Computing and Information Systems} \\
% \textit{University of Melbourne}\\
% Melbourne, Australia \\
% toby.murray@unimelb.edu.au}
% }

\maketitle

\begin{abstract}
    \noindent Neural networks are often susceptible to minor perturbations in input that cause them to misclassify.
    A recent solution to this problem is the use of globally-robust neural networks,
    which employ a function to certify that the classification of an input cannot be altered by such a perturbation.
    Outputs that pass this test are called \emph{certified robust}.
    However, to the authors' knowledge, these certification functions have not yet been verified at the implementation level.
    We demonstrate how previous unverified implementations are exploitably unsound in certain circumstances.
    Moreover, they often rely on approximation-based algorithms, such as power iteration, that (perhaps surprisingly) do not guarantee soundness.
    To provide assurance that a given output is robust, we implemented and formally verified a certification function for
    globally-robust neural networks in Dafny.
    We describe the program, its specifications,
    and the important design decisions taken for its implementation and verification, as well as our experience applying it in practice.
\end{abstract}

\section{Introduction}

Neural networks are deployed in safety- and security-critical
systems such as object recognition and malware classification. For these kinds
of models, it is important to be able to trust their outputs.
One important guarantee is that of \emph{robustness}, motivated by the existence of \emph{adversarial examples}~\cite{szegedy2013intriguing}: that a small change to
the model's input would not have caused it to produce a different output.

No useful classifier can be robust everywhere. For this reason,
a common approach is to assure the robustness of individual model outputs.
Certified robustness is a prominent approach for doing so, and
includes techniques like randomised smoothing~\cite{cohen2019},  
those based on enforcing differential privacy~\cite{lecuyer2019},
and those that leverage the certification procedure during model
training~\cite{lee2020}.

Many of these methods come with pen-and-paper proofs of their soundness:
theorems that provide a degree of confidence that an output will be robust if it is certified as robust.
However, for high-assurance applications of neural networks, 
pen-and-paper proofs fall short of the kinds of guarantees enjoyed by
\emph{formally verified}
safety- and security-critical systems software. For example, separation
kernels~\cite{murray2013} and cryptographic implementations~\cite{chudnov2018} enjoy
\emph{mechanised} proofs about their \emph{implementations}.

So-called ``code-level'' guarantees are important to rule out both
design- and implementation-level flaws that might otherwise compromise
robustness.

This paper considers the question of how to provide formally verified
guarantees of certified robustness. Doing so requires being able to overcome
key challenges. The first challenge is that work on formally verified robustness
considers primarily \emph{local robustness}, which means that it can require
symbolic reasoning to be performed for each output point or each perturbation bound that is to be
certified as robust~\cite{meng2022}.  This limits the efficiency of output certification. The second challenge is the complexity of many local
robustness verification or certification approaches. 
Formally verifying their implementations is prohibitive, as the effort required
to formally verify a program's implementation is known to be about an order
of magnitude higher than that to program it~\cite{klein2014}.

We overcome both of these challenges by designing a formally verified
robustness certifier for dense ReLU neural networks, which is inspired by Leino et al.'s training method for globally robust neural networks~\cite{leino2021}. In their work, a model's Lipschitz constants are estimated during training and used to maximise the model's robustness against
the training set. These constants are also used at inference time to cheaply certify
the robustness of individual outputs. We adapt this design to produce a formally verified robustness certifier that works in two stages: the first stage verifiably pre-computes Lipschitz upper bounds once-and-for-all, while the second stage is then executed for each model output to verifiably check its robustness against the Lipschitz upper bounds. Both stages have been formally verified in the industrial program verifier Dafny~\cite{leino2010}.

Along the way we uncovered soundness issues in the design (\cref{sec:power}) and implementation (\cref{sec:vuln}) of previous certified global robustness certifiers. We overcome these problems by adopting state-of-the-art algorithms for computing Lipschitz bounds~\cite{delattre2023}, which we implement and formally verify, along with our certifier routine. In addition, because our verification applies to the code of our certifier, it rules out soundness issues caused by floating point rounding (see~\cref{sec:vuln-fp}) in the certification function~\cite{jin2024getting}. Some orthogonal residual floating-point issues remain with our certifier, which we describe later in \cref{sec:limitations}.

This paper makes the following contributions:
\begin{itemize}
\item We design a verifiable certifier for robustness based on globally robust neural networks proposed by Leino et al.~\cite{leino2021},
\item We formalise its soundness as Dafny specifications for the corresponding top-level functions,
\item We present the implementation of our design, including how it overcomes the soundness issues explained above,
\item We formally verify our implementation in Dafny against its soundness specifications, obtaining a usefully applicable executable implementation.
\end{itemize}

We present a high-level overview of this paper's
main contributions in \cref{sec:overview}. The top-level specifications of
soundness we describe in \cref{sec:spec}. Key aspects of the implementation
we discuss in \cref{sec:cert} (the certification procedure), \cref{sec:norms} (deriving operator norms), and \cref{sec:roots} (positive square roots). \cref{sec:eval} reports on our experience applying our certifier to practical globally-robust image classification models whose size is on par with recent work on verification of global robustness properties~\cite{kabaha2024verification}. In \cref{sec:related-work}, we consider our approach in relation to prior work and conclude. \ifExtended \else Some technical details is relegated to the appendices contained in the extended version of this paper~\cite{EXTENDED}. \fi
%In \cref{sec:limitations}, we conclude by discussing the limitations of our current implementation and verification and avenues for future work.

\iffalse % old bullet list before first draft of intro
\begin{itemize}
	\item Neural networks are vulnerable to adversarial inputs.
	\item This is bad because [reasons].
	\item It is impossible for a neural network to be robust for all inputs.
	\item A well-known solution to this problem is certified robustness.
	\item There are many [techniques] for certifying robustness.
	\item These techniques certify local robustness.
	\item Some of these have even been verified to provide assurance.
	\item More recently, globally-robust neural networks have been developed.
	\item These work by pre-computing Lipschitz bounds.
	\item In addition to efficiency, these provide a metric of general robustness.
	\item No one yet has verified an implementation of global robustness certification.
	\item In this work, we implemented and verified global robustness certification in Dafny.
	\item This required tweaks to the original design, which we discovered was unsound.
	\item Our paper describes the unique design decisions we made, the specification of the Dafny program, and the proofs that were implemented.
\end{itemize}
\fi

\section{Exploitable Vulnerabilities in a Robustness Certifier}\label{sec:vuln}

We further motivate our work by describing a series of exploitable
vulnerabilities we
discovered in the robustness certifier implementation of Leino et al.~\cite{leino2021}. All are ruled out by
our verification. Two are implementation flaws (i.e.,\ bugs) that we reported
to the developers. 
The third results inevitably from the use of floating point arithmetic in their implementation, which our certifier
eschews.

\subsection{Incorrect Lipschitz Constant Computation}
\label{sec:vuln-normalisation}

The first vulnerability arises due to a subtle bug in the
implementation of the routine that calculates Lipschitz constants in
Leino et al.'s
implementation\footnote{\url{https://github.com/klasleino/gloro/issues/8}}.
An adversary who is able to choose a model's initial weights can cause
their certifier to incorrectly classify non-robust points as robust,
even after robust model training from those initial weights.

This issue could be exploited, for example, by an adversary who posts
a model online that purports to be accurate while enjoying a certain
level of robustness. Anyone who attempts to use that model in
conjunction with Leino et al.'s certifier can be mislead into
believing the model really is as robust as it purports to be, when in
fact its true robustness can be dramatically lower.

This vulnerability arises for models with small weights.
To exploit it we trained an ordinary (non-robust) MNIST
model, achieving an accuracy of 98.45\%. We then repeatedly halved all weights
in the second-to-last model layer while also doubling all weights in the model's
final layer. Doing so does not meaningfully change the model's Lipschitz
constants.
However, after repeating this process for a number of iterations, Leino et al.'s
implementation mistakenly computes very small, misleading Lipschitz constants
that then cause it to mistakenly certify non-robust outputs as robust, even for
large perturbation bounds~$\epsilon = 1.58$. When evaluating this model on the 10,000 MNIST
test points, their certifier mistakenly reports a \emph{Verified Robust Accuracy} (VRA)
measure~\cite{leino2021} of 98.42\%. VRA is the percentage of points that the model
accurately classifies and that their certifier says are robust at $\epsilon=1.58$.
\iffalse % cut for space
This figure is well above
the best reported VRA for MNIST at $\epsilon=1.58$, which is 62.8\%~\cite{leino2021} and
indeed it is wrong.
\fi
We were able to generate adversarial examples at $\epsilon=1.58$
for 8,682 of the test points that Leino et al.'s certifier said were robust.
\iffalse % speaks for itself?
One might conclude
therefore that their certifier was unsound at least 86.82\% of the time when applied to this model.
\fi
%
Further information about this issue is in \ifExtended\cref{app:unsound-normalization}\else{the extended version~\cite{EXTENDED}}\fi.

\subsection{Incorrect Certification}
\label{sec:vuln-certification}

The second vulnerability results from a subtle error in their certification routine that,
given the computed Lipschitz constants, certifies individual outputs\footnote{\url{https://github.com/klasleino/gloro/issues/9}}. An adversary who is able to supply specially crafted inputs to a model can cause Leino et al.'s certifier
to mistakenly certify the model's output as robust.

This issue causes Leino et al.'s certifier to certify as robust any output vector whose individual
elements are all equal. We validated that it can be exploited by an adversary who has white-box
access to a model (i.e.,\ knows the model architecture and weights). The adversary can simply perform a
gradient descent search to find any input that produces the \textbf{0} output vector (whose elements are all zero),
e.g., by using the mean absolute error (MAE) between the model's output and the target output vector~\textbf{0} as
the loss function.

We confirmed that this approach is able to find inputs that Leino et al.'s implementation will mistakenly certify
as robust at \emph{any} perturbation bound~$\epsilon$ for the MNIST, Fashion MNIST, and CIFAR-10 models considered
in this paper (see \cref{sec:eval}).
Further information about this issue is in \ifExtended\cref{app:unsound-neginf}\else{the extended version~\cite{EXTENDED}}\fi.

\subsection{Floating Point Imprecision}
\label{sec:vuln-fp}

The final issue arises due to floating point imprecision in Leino et al.'s implementation. This causes it to mistakenly
compute Lipschitz constants of 0 for models with very tiny weights. It can be exploited using the same method
as the issue in~\cref{sec:vuln-normalisation}. Further information about this issue is in \ifExtended\cref{app:unsound-fp}\else{\cite{EXTENDED}}\fi.

\section{Overview}\label{sec:overview}

\begin{figure}[t]
    \centering
    \includegraphics[width=0.8\linewidth]{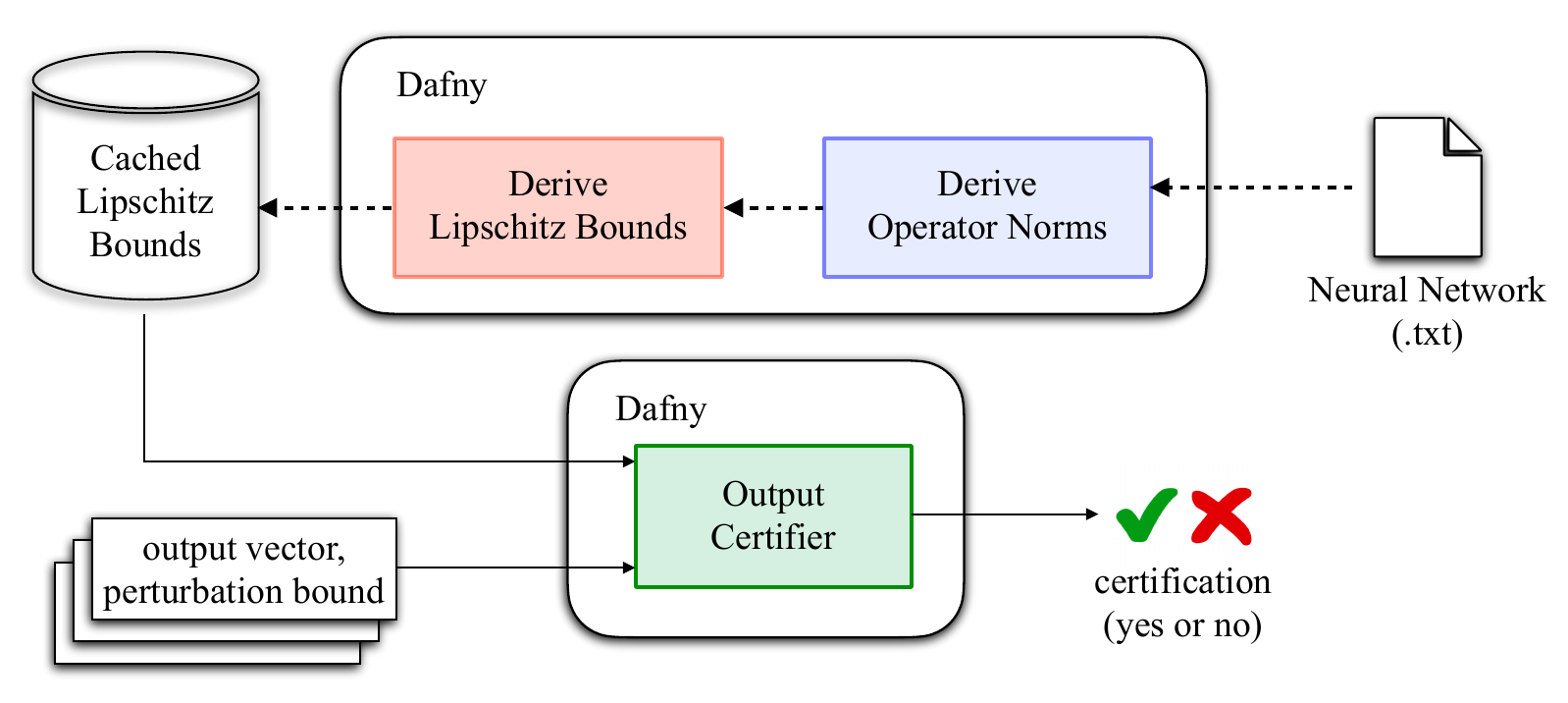}
\caption{An overview of the Dafny program. Lipschitz bounds are pre-computed and then reused as each model output is certified against a given perturbation bound.}
\label{overview}
\end{figure}
\iffalse
\begin{figure*}
\centering
\tikzstyle{process} = [rectangle, minimum height=1cm, text centered, draw=black, fill=orange!30]
\tikzstyle{input} = [circle, minimum width=0.3cm, draw=blue]
\tikzstyle{arrow} = [thick,->,>=stealth]
\tikzstyle{data} = [thick,dashed,->,>=stealth,blue]
\begin{tikzpicture}[node distance=4cm]
  \node (nn) [input] {};
  \node (sn) [process, below of=nn, yshift=2cm] {Derive operator norms};
  \node (lb) [process, below of=sn, yshift=2cm] {Derive Lipschitz bounds};
  \node (lb_data) [cylinder, draw, minimum height=1cm, minimum width=1cm, shape aspect=0.5, shape border rotate=90, cylinder uses custom fill, cylinder body fill=blue!20, cylinder end fill=blue!40, right of=lb, xshift=1cm] {};
  \node (cert) [process, below of=lb, yshift=2cm] {Certify output};
  \node (ov) [input, left of=cert] {};
  \node (out) [input, below of=cert, yshift=2cm] {};
  \draw [data] (nn) -- (sn) node[midway, left] {Neural network (.txt)};
  \draw [arrow] (sn) -- (lb);
  \draw [data] (lb) -- (lb_data) node[midway, above, align=left] {Lipschitz\\bounds};
  \draw [arrow] (lb) -- (cert);
  \draw [data] (lb_data) |- (cert);
  \draw [data] (ov) -- (cert) node[midway, above, align=left] {Output vector,\\perturbation bound};
  \draw [data] (cert) -- (out) node[midway, left] {Certification};
  \draw[arrow, looseness=3, out=295, in=340] (cert) to (cert);
\end{tikzpicture}
\caption{An overview of the Dafny program. Lipschitz bounds are pre-computed and then reused as each model output is certified.}
\label{overview}
\end{figure*}
\fi

\subsection{Robustness, Formally}

For our purposes, a neural network $N$ is a sequence of layers, represented by matrices $M_1,...,M_n$. Each layer $M_i$ can be \emph{applied} to a vector $\vv_i$ by taking their matrix-vector product $M_i\vv_i$. For all layers but the output layer~$M_n$, the ReLU activation function $R$ is then applied component-wise to the resulting vector, which we denote $R(M_i\vv_i)$. To apply a neural network $N$ with matrices $M_1,...,M_n$ to an input vector $\vv_1$, we simply apply each layer in-turn:
\begin{align*}
  N(\vv_1)&=M_n\vv_n\ \text{where}\\
  \qquad \vv_{i+1}&=R(M_i\vv_i) \text{ for } 1 \leq i < n.
\end{align*}
Let $\vv[i]$ denote the $i$th component of $\vv$. Formally, the output vector $N(\vv)$ is \emph{robust} with respect to some perturbation bound $\epsilon$ if:
\begin{equation}\label{eqn:robust}
  \forall\uu:||\vv-\uu||\le\epsilon\implies\func{ArgMax}(N(\vv))=\func{ArgMax}(N(\uu))
\end{equation}
where $||\cdot||$ denotes the $l_2$ norm:
$||\vv||=\sqrt{\sum_{i=1}^{|\vv|}\vv[i]^2}$. This condition guarantees that $\vv$ is not an $\epsilon$~adversarial example~\cite{mangal2023certifying}.

\subsection{The Global-Robustness Approach}\label{sec:margin-mention}

Figure~\ref{overview} illustrates our Dafny implementation of the global-robustness approach developed by Leino et al.~\cite{leino2021} for certifying this condition. Distinctively, the approach involves deriving and caching \emph{margin Lipschitz bounds} which can then be leveraged to efficiently certify any output vector against any perturbation bound. In our Dafny implementation, we generate margin Lipschitz bounds $L_{i,j}$ for distinct dimensions $i,j$ of the neural network's output vector. 
These can be best described as upper bounds on the rate at which the difference between these two dimensions can change, relative to changes in the input vector.
Formally, for a neural network $N$:
\[
  \forall\vv,\uu:\frac{|N(\vv)[j]-N(\vv)[i]-(N(\uu)[j]-N(\uu)[i])|}{||\vv-\uu||}\le L_{i,j}
\]
From this, follows: 
\[
  \forall\vv,\uu,e:||\vv-\uu||\le e\implies |N(\vv)[j]-N(\vv)[i]-(N(\uu)[j]-N(\uu)[i])| \le eL_{i,j}
\]
That is, $eL_{i,j}$ bounds the change in the difference between each pair of distinct components $i,j$ in the output vector. Given this fact, we can prove the consequent of \cref{eqn:robust} by considering the maximum component $j=\func{ArgMax}(N(\vv))$ and checking that the difference between it and each other component~$i \ne j$ is greater than~$eL_{i,j}$. Formally, we need to check that:
\[
  \forall i\ne j: N(\vv)[j]-N(\vv)[i] > eL_{i,j}
\]
for each other component $i$ in the output vector.

An important advantage of precomputing Lipschitz bounds is that they not only provide a way to efficiently certify outputs, but they also represent a global metric for the general robustness of the neural network. That is, neural networks with smaller Lipschitz bounds are robust for a broader range of outputs.
%It is for this reason that a core feature of Leino et al.'s certified global robustness~\cite{leino2021} is a training method for minimising these bounds.

During training, Leino et al.~\cite{leino2021} use an efficient method for estimating Lipschitz bounds to incorporate into the training objective the maximisation of the model's robustness against the training set. Unfortunately, the method that Leino et al. use for computing Lipschitz bounds is not guaranteed to be sound and is thus unsuitable to be implemented and verified in Dafny. Instead, we take advantage of the fact that our Dafny program is used after training has been completed, to verifiably pre-compute sound Lipschitz bounds for later use during output certification (i.e.\ that are used later at inference time). Therefore we can employ a sound but less efficient method to compute these bounds.

\subsection{Deriving Lipschitz Bounds}

To derive the Lipschitz bounds of our neural network, we first derive upper bounds on the operator norms of the first $n-1$ matrices. These can be thought of as Lipschitz bounds over the output vectors of the respective layers. Formally, the operator norm $||M||_{op}$ of matrix $M$ satisfies by definition:
$\forall\vv:\frac{||M\vv||}{||\vv||}\le||M||_{op}$.
By replacing $\vv$ with $\vv-\uu$ and distributing the matrix-vector product, we observe that $||M||_{op}$ is a Lipschitz bound on multiplication by $M$:
\begin{equation}\label{eqn:two}
  \forall\vv,\uu:\frac{||M\vv-M\uu||}{||\vv-\uu||}\le||M||_{op}
\end{equation}

The final step of applying a (non-output) layer is the component-wise application of the ReLU function $R$ to the resulting vector. The function applied to each component is:
$R(x)\sdef\func{max}(0,x)$.
Now note that for any two inputs to this function, the absolute difference in the outputs is less than or equal to the absolute difference in the inputs. Formally:
$\forall x,y:|R(x)-R(y)|\le|x-y|$.
Hence, for any two vectors $\ww,\xx$ of equal length, the absolute difference between each component $|\ww_i-\xx_i|$ is the same or less after applying $R$ to $\ww_i$ and $\xx_i$. Formally, by replacing $x$ in the above with $\ww_i$ and $y$ with $\xx_i$:
\[
  \forall i:|R(\ww_i)-R(\xx_i)|\le|\ww_i-\xx_i|
\]
Therefore, by applying $R$ to each component in $\ww,\xx$, we reduce the distance in each dimension of the vector. We therefore decrease the distance overall:
$||R(\ww)-R(\xx)||\le||\ww-\xx||$.
Replacing $\ww$ and $\xx$ with $M\vv$ and $M\uu$:
$||R(M\vv)-R(M\uu)||\le||M\vv-M\uu||$.
And therefore, from \cref{eqn:two}:
\[
  \forall\vv,\uu:\frac{||R(M\vv)-R(M\uu)||}{||\vv-\uu||}\le||M||_{op}
\]
Hence, the operator norm of a matrix $M$ is a Lipschitz bound on the application of a layer represented by $M$.

A Lipschitz bound over the output of the first $n-1$ layers of a neural network can therefore be derived as the product of their operator norms:
\begin{equation}\label{eqn:product}
  \prod_{i=1}^{n-1}||M_i||_{op}.
\end{equation}

When computing the output of neural network $N$, the value of the $i$th component of the output vector $\vv_{n+1}$ is equal to the dot product of the input vector~$\vv_n$ to final matrix $M_n$, with the $i$th row in $M_n$. Hence, a margin Lipschitz bound on the difference between the $j$th and $i$th components can be derived by multiplying the product~\cref{eqn:product} by the operator norm of $M_n[j]-M_n[i]$ (where for matrix~$M$ we write $M[k]$ denote its $kth$ row, indexed from 0). Because
$M_n[j]-M_n[i]$ is a single vector,
we prove in Dafny that (due to the Cauchy-Schwartz inequality) its operator norm can be efficiently bounded by its $l_2$ norm: $||M_n[j]-M_n[i]||$.

\section{Top-Level Specification}\label{sec:spec}

This section details our encoding of the robustness condition in Dafny and the specifications our robustness certifier is verified against.

\subsection{Types}

For a type $\mathbf{t}$, the type $\Seq{\mathbf{t}}$ is the type of sequences of $\mathbf{t}$. Given a sequence $x$, we write $x[i]$ to denote the $i$th element of~$x$ (indexed from 0).

We define the type $\Vector$ to be a non-empty sequence of $\Real$s. These $\Real$s Dafny compiles to arbitrary-precision rationals, and ensure
our certifier avoids the floating-point unsoundness issues of Leino et al.'s implementation that we identify in \cref{sec:vuln-fp}.
\begin{flalign*}
&\Vector=\{v\in\Seq{\Real}\mid|v|>0\}&
\end{flalign*}
A matrix is a non-empty sequence of vectors with equal dimension:
\begin{flalign*}
&\Matrix=\{M\in\Seq{\Vector}\mid|M|>0&\\
&\quad\land\forall i,j\in\Nat\:.\:i<|M|\land j<|M|\implies|M[i]|=|M[j]|\}
\end{flalign*}
We define each vector in a matrix to be a \emph{row} of that matrix. We can define Dafny functions that return the number of rows and columns in a matrix:
\begin{tabular}{ll}
  \begin{minipage}{0.45\textwidth}
\begin{flalign*}
\TotalSpec
  {\FUNCTION
    {\Rows}
    {M: \Matrix}
    {\Nat}
  }
  {|M|}
\end{flalign*}
  \end{minipage} &
  \begin{minipage}{0.45\textwidth}
\begin{flalign*}
\TotalSpec
  {\FUNCTION
    {\Cols}
    {M: \Matrix}
    {\Nat}
  }
  {|M[0]|}
\end{flalign*}
\end{minipage}
\end{tabular}
\medskip

A neural network is a non-empty sequence of matrices where the number of rows in each matrix is equal to the number of the columns in the next:
\begin{flalign*}
&\NeuralNetwork=\{N\in\Seq{\Matrix}\mid|N|>0&\\
&\quad\land\forall i\in\Nat\:.\:i<|N|-1\implies\Rows(N[i])=\Cols(N[i+1])\}
\end{flalign*}
A vector is a compatible input to a neural network if and only if its dimension is compatible with multiplication by the first matrix:
\begin{flalign*}
\TotalSpec
  {\FUNCTION
    {\CompatibleInput}
    {v:\Vector,\ N:\NeuralNetwork}
    {\Bool}
  }
  {|v|=\Cols(N[0])}
\end{flalign*}
Similarly, $v$ is a compatible output of $N$ if and only if its dimension is equal to that of the matrix-vector product of the final matrix and its input vector:
\begin{flalign*}
\TotalSpec
  {\FUNCTION
    {\CompatibleOutput}
    {v:\Vector,\ N:\NeuralNetwork}
    {\Bool}
  }
  {|v|=\Rows(N[|N|-1])}
\end{flalign*}

\subsection{Modelling the Neural Network}
\label{sec:neural-net-model}

The application of a neural network is modelled with the function $\ApplyNeuralNet$, which makes use of the recursive
helper $\ApplyNeuralNetBody$ that model the application of all layers but the final one:
\begin{flalign*}
\Spec
  {\CompatibleInput(v,N)}
  {\FUNCTION
    {\ApplyNeuralNetBody}
    {N:\NeuralNetwork,\ v:\Vector}
    {\Vector}
  }
  {\\&\quad\ITE{|N|=1}{\ApplyLayer(N[0],v)\\&\quad}{\ApplyLayer(N[|N|-1],\ApplyNeuralNetBody(N[..|N|-1],v))}}
\end{flalign*}
\begin{flalign*}
\Spec
  {\CompatibleInput(v,N)}
  {\FUNCTION
    {\ApplyNeuralNet}
    {N:\NeuralNetwork,\ v:\Vector}
    {\Vector}
  }
  {\\&\quad\ITE{|N|=1}{\MV(N[0],v)\\&\quad}{\MV(N[|N|-1],\ApplyNeuralNetBody(N[..|N|-1],v))}}
\end{flalign*}
The matrix-vector product $\MV$ is defined in \ifExtended\cref{sec:appendix-matrix-ops}\else{the extended version~\cite{EXTENDED}}\fi.

The application of a non-final layer $\ApplyLayer$ is defined to use the ReLU activation function.
Note that the Dafny syntax $[x]$ for some variable $x$ denotes a sequence only containing $x$, and that + is sequence concatenation.
\begin{flalign*}
\Spec
  {|v|=\Cols(M)}
  {\FUNCTION
    {\ApplyLayer}
    {M:\Matrix,\ v:\Vector}
    {\Vector}
  }
  {\\&\quad\ApplyRelu(\MV(M,v))}
\end{flalign*}
\begin{flalign*}
\TotalSpec
  {\FUNCTION
    {\ApplyRelu}
    {v:\Vector}
    {\Vector}
  }
  {\Apply(v, Relu)}
\end{flalign*}
\begin{flalign*}
\TotalSpec
  {\FUNCTION
    {\Apply}
    {v:\Vector,\ f:\Real\rightarrow\Real}
    {\Vector}
  }
  {\\&\quad\ITE{|v|=1}{[f(v[0])]\ }{[f(v[0])]+\Apply(v[1..],f)}}
\end{flalign*}
\begin{flalign*}
\TotalSpec
  {\FUNCTION
    {\Relu}
    {x:\Real}
    {\Real}
  }
  {\ITE{x\ge0}{x\ }{0}}
\end{flalign*}

\subsection{Linear Algebra}

Encoding the $l_2$ norm in Dafny requires building up a small set of basic mathematical functions:
\begin{flalign*}
\TotalSpec
  {\FUNCTION
    {\LTwo}
    {v:\Vector}
    {\Real}
  }
  {\Sqrt(\Sum(\Apply(v,\Square)))}
\end{flalign*}
The (positive) square root function cannot be defined directly and is therefore not compilable. However its properties can still be specified in Dafny. We do so by specifying it as a \emph{ghost function} with postcondition (``$\kw{ensures}$") annotations that precisely describe what it means for a real $r$ to be the square root of a non-negative real $x$:
\begin{flalign*}
\GhostSpec
  {x\ge0}
  {\GHOST
    {\Sqrt}
    {x:\Real}
    {(r:\Real)}
  }
  {r\ge0\land r\cdot r=x}
\end{flalign*}
The $\Sum$ and $\Square$ functions are straightforward (see \ifExtended\cref{sec:appendix-matrix-ops}\else{the extended version~\cite{EXTENDED}}\fi).

%This suffices for the definition of $\LTwo$.
For our definition of the robustness property, we additionally need to define vector subtraction and the $\ArgMax$ function:
\begin{flalign*}
\Spec
  {|v|=|u|}
  {\FUNCTION
    {\Minus}
    {v:\Vector,\ u:\Vector}
    {\Vector}
  }
  {\\&\quad\ITE{|v|=1}{[v[0]-u[0]]\ }{[v[0]-u[0]]+\Minus(v[1..],u[1..])}}
\end{flalign*}
\begin{flalign*}
\TotalSpec
  {\FUNCTION
    {\ArgMax}
    {s:\Vector}
    {\Nat}
  }
  {\\&\quad\ITE{|s|=1}{0\\&\quad}
  {\ITE{s[\ArgMax(s[..|s|-1])]\ge s[|s|-1]}{\ArgMax(s[..|s|-1])\\&\quad}{|s|-1}}}
\end{flalign*}
Finally, for our convenience, we define a function that represents the distance between two vectors:
\begin{flalign*}
\Spec
  {|v|=|u|}
  {\FUNCTION
    {\Distance}
    {v:\Vector,\ u:\Vector}
    {\Real}
  }
  {\LTwo(\Minus(v,u))}
\end{flalign*}

\subsection{Robustness definition}

We can now define robustness in Dafny. For a given input vector $v$ with output vector $v'=\ApplyNeuralNet(N,v)$, we say $v'$ is robust with respect to perturbation bound $e$ if $\Robust(v,v',e,N)=\True$, where:
\begin{flalign*}
\Spec
  {\CompatibleInput(v,n)\land\ApplyNeuralNet(N,v)=v'}
  {\FUNCTION
    {\Robust}
    {v:\Vector,\ v':\Vector,\ e:\Real,N:\NeuralNetwork}
    {\Bool}
  }
  {\\&\quad\forall u\in\Vector\:.\:|v|=|u|\land\Distance(v,u)\le e
  \\&\quad\implies\ArgMax(v')=\ArgMax(\ApplyNeuralNet(N,u))}
\end{flalign*}
With global robustness certification, we can establish the robustness of an output vector irrespective of its input vector. That is, given
%that the user inputs
an output vector $v'$, a neural network $N$, and a perturbation bound $e$, our verified certifier says ``Certified" only if Dafny can verify the assertion:
\begin{flalign*}
&\Assert\ \forall v\in\Vector\:.\:\CompatibleInput(v,N)\land\ApplyNeuralNet(N,v)=v'&\\&\implies\Robust(v,v',e,N)
\end{flalign*}

\section{Verified Certification Procedure}\label{sec:cert}

As discussed in \cref{sec:overview}, for a neural network $N$ composed of $n$ matrices, a margin Lipschitz bound for the $i,k$th component-pair of the output vector can be derived by taking the product of the operator norms of the first $n-1$ matrices, together with the operator norm of the difference between the $k$th and~$i$th rows of the final matrix (bounded by its $l_2$ norm). We implement this computation in the Dafny method $\GenerateLipschitzBound$ in \cref{fig:dafny-GenerateLipschitzBound}, which generates a Lipschitz bound for the $i,k$th component-pair in the output vector of a neural network $N$, given a sequence $s$ containing the operator norms of all matrices in $N$. The conditions specified in the $\kw{requires}$ and $\kw{ensures}$ clauses state the precondition and postcondition of this method respectively.
\begin{figure}[t]
\begin{flalign*}
&\kw{method}\ \GenerateLipschitzBound(N: \NeuralNetwork,\ i:\Nat,\ k:\Nat,\ s:\Seq{\Real}):(r:\Real)&\\
&\quad\kw{requires}\ |s|=|N|\land i<\Rows(N[|N|-1]) \land k<\Rows(N[|N|-1]) \land i \not= k\\
&\quad\kw{requires}\ \forall j\in\Nat\:.\:j<|s|\implies s[j]\ge\OpNorm(N[j])\\
&\quad\kw{ensures}\ \IsMarginLipschitzBound(N, r, i, k)\\
&\{\\
&\quad\kw{var}\ i:=|N|-1\\
&\quad\kw{var}\ d:=\Minus(N[|N|-1][k],N[|N|-1][i])\\
&\quad\kw{var}\ r:= \LTwoUpperBound(d)\\
&\quad\kw{while}\ i > 0\\
&\quad\quad\kw{invariant}\ r \geq 0 \land \IsMarginLipschitzBound(N[i..], r, i, k)\\
&\quad\{\\
&\quad\quad i := i - 1\\
&\quad\quad r := s[i] * r\\  
&\quad\}\\
&\}
\end{flalign*}
\caption{Generating margin Lipschitz bounds in Dafny.\label{fig:dafny-GenerateLipschitzBound}}
\end{figure}

\begin{flalign*}
\Spec
  {i < |N[|N|-1]| \land k < |N[|N|-1]|}
  {\FUNCTION
    {\IsMarginLipschitzBound}
    {N:\NeuralNetwork, r:\Real, i:\Nat, k:\Nat}
    {\Bool}
  }
  {\\
&\forall v\in\Vector,\ u\in\Vector\:.\:\CompatibleInput(v,N)\land\CompatibleInput(u,N)\\
&\quad\quad\implies\Abs(\,\ApplyNeuralNet(N,v)[k]-\ApplyNeuralNet(N,v)[i]-\\
&\quad\quad\quad\qquad\qquad(\ApplyNeuralNet(N,u)[k]-\ApplyNeuralNet(N,u)[i])\,)\\
&\quad\quad\quad\le r\cdot\Distance(v,u)
   }
\end{flalign*}

The procedure begins by extracting the vector subtraction of the $k$th and $i$th rows of the final matrix in $N$ and storing this vector difference in $d$. The upper bound of the $l_2$ norm of this new vector is then assigned to $r$. The final Lipschitz bound is then derived by taking the product of the first $|s|-1$ elements of $s$, multiplied by $r$. The $\kw{invariant}$ annotation specifies while-loop's invariant.

The upper bound of the $l_2$ norm is computed by summing the squares of each vector element and then taking an upper bound of the square root
(see \cref{sec:roots} later). The operator norms in~$s$ are approximated using an iterative method described in \cref{sec:norms}.

To enable Dafny to verify the $\GenerateLipschitzBound$ method, we must first prove three facts. Firstly, that
an upper bound of the $l_2$ norm of a vector is also an upper bound on the operator norm of the matrix that comprises just that vector.
Secondly, that the operator norm bound of $M[k] - M[i]$ yields the margin Lipschitz bound for $i,k$ for a single-layer neural network.
These two facts establish that the
invariant holds when the while-loop is entered. Thirdly, to prove that the loop's invariant is maintained, we must show that
multiplying the margin Lipschitz bound by the operator norm bound for the matrix of the preceding layer yields the margin Lipschitz bound for the
composition of that preceding layer and the subsequent part of the neural network.  In Dafny, these facts are stated as \emph{lemmas}, which are uncompiled Dafny methods wherein the $\kw{ensures}$ clause is verified against the $\kw{requires}$ clause with a proof in the method body.
Essentially, the \kw{requires} clauses state the lemma's assumptions and the \kw{ensures} clause states its conclusion.
%Verifying the lemma proves that the conclusion can indeed be derived from the assumptions.
The three lemmas corresponding to these three facts appear in \cref{fig:dafny-lemmas-lipschitz}. The proof of the first leverages the
Cauchy-Schwartz inequality, which we axiomatise in Dafny. \label{sec:cauchy-schwartz}
% Toby: Actually the second lemma doesn't seem to correspond to the description of it above.
%       Also, perhaps the above needs to be extended or generalised for the margin Lipschitz bounds?

\begin{figure}[t]
\begin{flalign*}
&\kw{lemma}\ \mathit{L2IsOpNormUpperBound}(s:\Real,\ m:\Matrix)\\
&\quad\kw{requires}\ |m| = 1\\
&\quad\kw{requires}\ s \geq \LTwo(m[0])\\
&\quad\kw{ensures}\ s \geq \OpNorm(m)\\
\\[-1ex]
&\kw{lemma}\ \mathit{OpNormIsMarginLipBound}(N:\NeuralNetwork,\ m:\Matrix,\ \\
&\qquad\qquad\qquad\qquad\qquad\qquad\qquad\qquad\  i:\Nat,\ k:\Nat,\ r:\Real)\\
&\quad\kw{requires}\ |N| = 1\\
&\quad\kw{requires}\ i < |N[0]| \land k < |N[0]|\\
&\quad\kw{requires}\ m = [\Minus(N[0][k], N[0][i])]\\
&\quad\kw{requires}\ r \geq \OpNorm(m)\\
&\quad\kw{ensures}\ \IsMarginLipschitzBound(N, r, i, k)\\
\\[-1ex]
&\kw{lemma}\ \mathit{MarginRecursive}(N:\NeuralNetwork,\ s:\Real,\ r:\Real,\ i:\Nat,\ k:\Nat,\ \\
&\qquad\qquad\qquad\qquad\qquad\quad\ \  r':\Real)\\
&\quad\kw{requires}\ |N| > 1\\
&\quad\kw{requires}\ i < |N[|N|-1]| \land k < |N[|N|-1]|\\
&\quad\kw{requires}\ s \geq \OpNorm(N[0])\\
&\quad\kw{requires}\ \IsMarginLipschitzBound(N[1..], r, i, k)\\
&\quad\kw{requires}\ r' = s \cdot r\\
&\quad\kw{requires}\ r \geq 0\\
&\quad\kw{ensures}\ \IsMarginLipschitzBound(N, r', i, k)
\end{flalign*}
\caption{Lemmas used to prove \cref{fig:dafny-GenerateLipschitzBound}.\label{fig:dafny-lemmas-lipschitz}}
\end{figure}

With Lipschitz bounds generated and cached, the certification procedure is straightforward to implement and verify (though note that it is also
easy to introduce subtle bugs in these kinds of routines, as we found in Leino et al.'s implementation as described in~\cref{sec:vuln-certification}). Our certification method is shown in \cref{fig:dafny-Certify}, where $\AreLipBounds(N,L)$ specifies that each $L[i][k]$ is a margin Lipschitz bound for components~$i,k$ of~$M$, as specified by $\IsMarginLipschitzBound$ above.

\begin{figure}[t]
\begin{flalign*}
&\kw{method}\ \Certify(v':\Vector, e:\Real, L:\Seq{\Seq{\Real}}):(b:\Bool)&\\
&\quad\kw{ensures}\ b\implies \forall v\in\Vector,\ N\in\NeuralNetwork\:.\:\\
&\quad\quad\CompatibleInput(v,N)\land\ApplyNeuralNet(N,v)=v'\land\AreLipBounds(N,L)\\
&\quad\quad\implies\Robust(v,v',e,N)\\
&\{\\
&\quad\kw{var}\ x:=\ArgMax(v')\\
&\quad\kw{var}\ i:=0\\
&\quad b:=\True\\
&\quad\kw{while}\ i<|v'|\ \{\\
&\quad\quad\kw{if}\ i\not=x\ \{\\
&\quad\quad\quad\kw{if}\  L[i][x] \cdot e  \geq v'[x]-v'[i]\ \{\\
&\quad\quad\quad\quad b:=\False;\\
&\quad\quad\quad\quad\kw{break};\\
&\quad\quad\quad\}\\
&\quad\quad\}\\  
&\quad\quad i := i + 1;\\
&\quad\}\\
&\}
\end{flalign*}
\caption{Certification procedure implemented in Dafny.\label{fig:dafny-Certify}}
\end{figure}

As discussed in \cref{sec:overview}, this involves checking that for each other component $i$ in the output vector $v'$, the difference between the maximum value of~$v'$ and $v'[i]$ is less than the product of the corresponding margin Lipschitz bound with the perturbation bound $e$.

\section{Deriving Operator Norms}\label{sec:norms}

Matrix~$M$'s operator norm $||M||_{op}$ bounds how much it can ``stretch" a vector:
\[
  ||M||_{op}=\text{inf}\{c\ge 0\ \mid\ \forall\vv\:.\:||M\vv||\le c||\vv||\}
\]
In Dafny, we encode this definition as in \cref{fig:dafny-OpNorm}.
\begin{figure}[t]
\begin{flalign*}
\TotalGhostSpec
  {\GHOST
    {\OpNorm}
    {M:\Matrix}
    {(r:\Real)}
  }
  {r\ge0 \ \land \\&\quad
  (\forall v\in\Vector\:.\:|v|=\Cols(M) \implies\LTwo(\MV(M,v))\le r\cdot\LTwo(v))\  \land \\&\quad
  \neg\exists x\in\Real\:.\:0\le x<r\land\forall v\in\Vector\:.\:|v|=\Cols(M)\\&\quad\quad
  \implies\LTwo(\MV(M,v))\le x\cdot\LTwo(v)}
\end{flalign*}
\caption{Defining operator norms in Dafny.\label{fig:dafny-OpNorm}}
\end{figure}
Unfortunately, operator norms cannot be derived directly and must be computed with iterative approximation. In Leino et al.~\cite{leino2021}, operator norms are derived using the \emph{power method}~\cite{gouk2021}. For a given matrix $M$, this involves choosing a random initial vector $\vv_1$ and applying the recurrence:
$\vv_{i+1}=M^T M\vv_i$.
The operator norm can then be derived as
\begin{equation}\label{eqn:three}
	\frac{||M\vv_n||}{||\vv_n||}
\end{equation}
for some suitably large $n$. In practice, intermediary normalisation is performed for each $\vv_i$ to avoid overflow (though is easy to implement incorrectly; a bug here in Leino et al.'s implementation causes the
vulnerability of
\cref{sec:vuln-normalisation}).

Intuitively, this works because, as $i$ increases, the direction of $\vv_i$ converges to that of the maximum eigenvector of $M^T M$. This is the vector whose length is increased by the greatest factor when its product is taken with $M$.
\iffalse % cut for space, repetition
An approximation of the operator norm is derived as this factor in \cref{eqn:three}.
\fi

There are a number of issues with the power method that make it unsuitable for formal verification in Dafny. One issue is that, if the random initial vector is orthogonal to the maximum eigenvector of $M^T M$, the algorithm may fail to converge. Furthermore, the method converges on the operator norm from \emph{below}, since the result of the function in \cref{eqn:three} applied to intermediary values of $\vv_i$ is lower than that for the maximum eigenvector of $M^T M$, by definition.

\label{sec:power}
For these reasons, our Dafny implementation takes advantage of a relatively new approach for approximating operator norms from \emph{above}, called Gram iteration~\cite{delattre2023}. Unlike the power method, Gram iteration involves iterating on the matrix itself, rather than an initial starting vector. Let $M_0=M$ be the initial matrix. Gram iteration involves applying the recurrence:
\begin{equation}\label{eqn:four}
	M_{i+1}=M_i^T M_i
\end{equation}
Then, for some suitably-large $n$, we derive an upper bound on the operator norm as:
$\sqrt[2^n]{||M_n||_F}$,
where $||\cdot||_F$ is the Frobenius norm, defined as:% the square root of the sum of the squares of all elements in the matrix:
\[
||M||_F\sdef\sqrt{\sum_{i=1}^{|M|}\sum_{j=1}^{|M[0]|}M[i][j]^2}.
\]

This method relies on three key facts:
\begin{enumerate}[label={\bf F\arabic*.}]
	\item $\sqrt{||M^T M||_{op}}=||M||_{op}$.
	\item $||M||_{op}\le||M||_F$ for any real matrix $M$.
	\item As $i$ increases, $||M_i||_F$ approaches $||M_i||_{op}$.
\end{enumerate}
\newcommand{\Fact}[1]{\textbf{F{#1}}}
Gram iteration works by repeatedly taking the Gram matrix of $M$, as in \cref{eqn:four}, and then computing its Frobenius norm, which, due to fact~\Fact{2}, is an upper bound on its operator norm, but due to fact \Fact{3}, is a very close approximation. Due to fact \Fact{1}, we can then derive an upper bound on the operator norm of $M$ by taking the square root $n$ times. To enable verification, we encode facts \Fact{1} and \Fact{2} as axiomatic assumptions in Dafny.

\iffalse % cut for space
per \cref{fig:dafny-assumptions}.
\begin{figure}[t]
\begin{flalign*}
\TotalSpec
  {\AXIOM
    {\AssumptionOne}
    {M:\Matrix}
  }
  {\\&\quad\OpNorm(M)\le\Sqrt(\OpNorm(\MatrixMatrixProduct(\Transpose(M),M)))}
\end{flalign*}
\begin{flalign*}
\TotalSpec
  {\AXIOM
    {\AssumptionTwo}
    {M:\Matrix}
  }
  {\\&\quad\OpNorm(M)\le\FrobeniusNorm(M)}
\end{flalign*}
\caption{Verification assumptions (axioms) encoded in Dafny.\label{fig:dafny-assumptions}}
\end{figure}
\fi % cut for space the formal encoding of the assumptions

Naively applying recurrence~\cref{eqn:four} quickly leads to having to compute matrix multiplication on very large numbers. Therefore,
our implementation normalises the result on each iteration by dividing by the Frobenius norm and then truncating the result to 16
decimal places. Dividing by the Frobenius norm has a predictable impact on the matrix's operator norm, since $||M||_{op} \leq ||\frac{M}{x}||_{op} \cdot x$ for all $x > 0$. 
Truncation necessarily introduces errors into the resulting estimate of the matrix's operator norm. However, we can
track and bound the error introduced. For a matrix~$M$, let $\Truncate(M)$ denote its truncation and define $E = M - \Truncate(M)$ be the \emph{error}
introduced by truncation. Then, by Weyl's inequality, when~$M$ is a square, symmetric matrix (as all $M^T M$ are), we have
$|\:||M||_{op} - ||Truncate(M)||_{op}\:| \leq ||E||_{op}$ (since the operator norm is also the matrix's largest eigenvalue).
Thus $||M||_{op} \leq ||\Truncate(M)||_{op} + ||E||_{op}$. So, each Gram iteration computes
\[
M_{i+1}=\Truncate\left(\frac{M_i^T M_i}{||M_i^T M_i||_F}\right)
\]
and we have $||M_i||_{op} \leq \sqrt{r_{i+1} \cdot (||M_{i+1}||_{op} + ||E_{i+1}||_{op})}$ where
$r_{i+1} = ||M_i^T M_i||_F$ and $E_{i+1} = \left(\frac{M_i^T M_i}{||M_i^T M_i||_F}\right) - \Truncate\left(\frac{M_i^T M_i}{||M_i^T M_i||_F}\right)$.

Our verified Dafny implementation of Gram iteration appears in \cref{fig:dafny-GramIteration}.
\begin{figure}[t]
\begin{flalign*}
&\kw{method}\ \GramIteration(M:\Matrix,\ n:\Nat):(\mathit{ret}:\Real)&\\
&\quad\kw{ensures}\ \mathit{ret}\ge\OpNorm(M)\\
&\{\\
&\quad\kw{var}\ i:=0\\
&\quad\kw{var}\ a:=[]\\  
&\quad\kw{while}\ i\ne n\ \{\\
&\quad\quad M':= \mathit{MTM}(M)\\
&\quad\quad r:= \kw{if}\  \mathit{IsZeroMatrix}(M')\ \kw{then}\ 1\ \kw{else}\ \FrobeniusNormUpperBound(M')\\
&\quad\quad M,E:=\TruncateWithError(\MatrixDiv(M',r))\\
&\quad\quad a := [(r,\FrobeniusNormUpperBound(E)] + a\\
&\quad\quad i:=i+1\\
&\quad\}\\
&\quad \mathit{ret}:=\FrobeniusNormUpperBound(M)\\
&\quad \mathit{ret}:=\Expand(a,\mathit{ret})\\
&\}\\[2ex]
&\kw{function}\ \Expand(a:[(\Real,\Real)],\ v:\Real)\ \kw{returns}\ \Real&\\
&\quad \Expand([],\ v) = v&\\
&\quad \Expand((r,e):a,\ v) = \Expand(a,\ \SqrtUpperBound(r\cdot(v+e)))&\\  
\end{flalign*}
\caption{Gram iteration in Dafny.\label{fig:dafny-GramIteration} For an element $x$ and sequence~$xs$, we write $x:xs$ to denote
the sequence whose head is~$x$ and whose tail is~$xs$.}
\end{figure}
\label{sec:mtm}
This method accepts a matrix $M$ and a natural number $n$ which determines the number of iterations to apply. The return value $r$ is verified to be an upper bound on the operator norm $\OpNorm(M)$ as we have defined it in Dafny (as specified by the \textbf{ensures} postcondition annotation). For $n$ iterations, the algorithm repeatedly redefines $M$ to be its own Gram matrix, with normalisation and truncation as described above (taking care to avoid division by zero during normalisation).
The implementation uses a specialised, optimised routine~$\mathit{MTM(M)}$ for calculating $\MatrixMatrixProduct(\Transpose(M),M)$ that avoids explicitly
computing the transpose. This routine converts the matrix~$M$ to a two-dimensional array of $\Real$s for efficient access, transposing~$M$ during conversion.
Then each entry $M'[i][j]$ of the product $M'$ is the dot product of the two rows~$i$ and~$j$ of the transposed~$M$, and so can be efficiently
computed by row-wise scanning (maximising cache locality). Since the resulting~$M'$
is symmetric, this routine avoids calculating the lower diagonal by reusing the results from the upper diagonal.

Returning to \cref{fig:dafny-GramIteration}, the sequence~$a$ tracks quantities~$r$ and~$e$ corresponding to the scaling factor and error upper bound introduced by normalisation and truncation respectively.  The return-variable $\mathit{ret}$ is then set to a verified upper bound of the Frobenius norm of $M$, from which the verified upper bound of the operator norm is then computed by using the~$r$ and~$e$ terms to \emph{expand} this quantity, via the $\Expand$ function.
That function makes use of the $\SqrtUpperBound$ function for deriving upper bounds on square roots, discussed in the next section.

\section{Generating Positive Square Roots}\label{sec:roots}

To derive upper bounds on square roots we implement a version of Heron's method (aka the Babylonian method), shown in \cref{fig:dafny-SqrtUpperBound}. %The correctness proof is omitted for brevity.
This is an ancient algorithm for deriving square roots, whose correctness proof is relatively straightforward and is guaranteed to generate an upper bound.

\begin{figure}[tb]
\begin{flalign*}
&\kw{method}\ \SqrtUpperBound(x:\Real):(r:\Real)&\\
&\quad\kw{requires}\ x\ge 0\\
&\quad\kw{ensures}\ \Sqrt(x)\le r\\
&\{\\
&\quad\kw{if}\ x=0\ \{\\
&\quad\quad\kw{return}\ 0\\
&\quad\}\\ 
&\quad r:=\kw{if}\ x<1\ \kw{then}\ 1\ \kw{else}\ x\\
&\quad i:=0\\
&\quad\kw{while}\ i<\mathit{SQRT\_ITERATIONS}\ \{\\
&\quad\quad r_0:=r\\
&\quad\quad r:=(r+x/r)/2\\
&\quad\quad i:=i+1\\
&\quad\quad\kw{if}\ r_0-r\le \mathit{SQRT\_ERR}\ \{\\
&\quad\quad\quad\kw{return}\ r\\
&\quad\quad\}\\ 
&\quad\}\\
&\quad \kw{print}\text{ ``Warning: Sqrt algorithm terminated early."}\\
\}
\end{flalign*}
\caption{Heron's method for computing square root upper bounds in Dafny.\label{fig:dafny-SqrtUpperBound}}
\end{figure}

Our loop maintains the invariant $r\ge\Sqrt(x)$, which holds upon entry due to the preceding ternary assignment to $r$. We then iterate until the desired precision is attained, or the maximum number of iterations is reached. These parameters are encoded as the global constants $\mathit{SQRT\_ERR}$ and $\mathit{SQRT\_ITERATIONS}$ (in our current implementation $10^{-11}$ and $2\times10^6$ respectively). 

\iffalse % cut for space
We choose to use $\mathit{SQRT\_ITERATIONS}$ to bound the number of loop iterations to make
the required termination proof obvious. 
\fi

\section{Applying the Certifier}\label{sec:eval}

\begin{figure}[t]
  %\begin{tabular}{l@{\qquad}r}
  %\includegraphics[width=0.45\linewidth]{performance_7.pdf} &  \includegraphics[width=0.45\linewidth]{performance_14.pdf}
  \centering\includegraphics[width=0.85\linewidth]{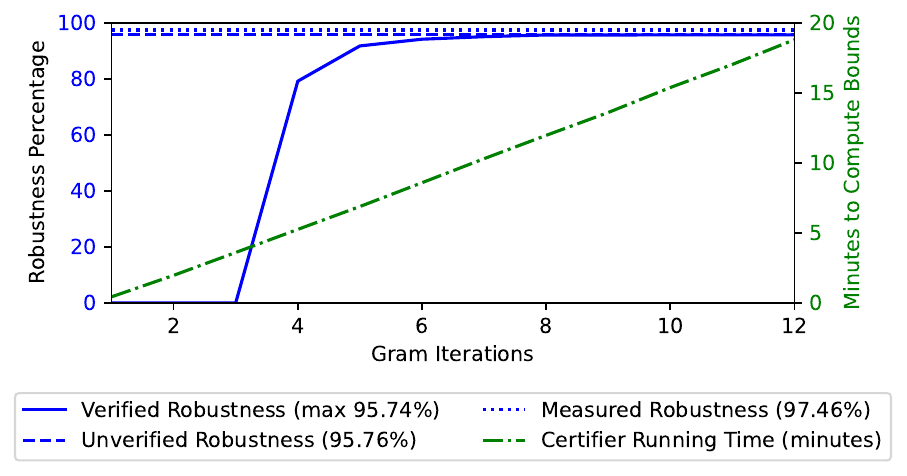}
  %\end{tabular}
  \caption{Certifier performance on globally robust MNIST model ($\epsilon = 0.3$).\label{fig:performance}}
\end{figure}

\iffalse % cut for space
Our certifier's implementation and verification (specifications, lemmas, etc.) together constitute 3.5K lines of Dafny.
\fi

After neural network training, our certifier is applied to compute safe Lipschitz bounds once-and-for-all. It is then repeatedly applied, having
computed those bounds, to certify individual model outputs.  The
time required to certify each output vector~$v'$ 
is linear in the vector's length~$|v'|$ and cheap: in the worst case it requires
less than $2|v'|$ loads, $5|v'|$ comparisons, and $|v'|$ negations, subtractions, and multiplications each (\cref{fig:dafny-Certify}), where each of these operations is performed over arbitrary-precision rationals (to which Dafny's $\Real$s are compiled). In practice, this means each individual output point requires approx.\ 8~milliseconds to certify (including text parsing, I/O and printing), \emph{independent of the model size}. Alternative approaches report median certification times per individual output of
anywhere from 10~milliseconds to 7.3~seconds~\cite[Table 2]{DBLP:conf/iclr/FromherzLFPP21} on comparable models to those that we consider below.

Therefore, 
we seek to understand~(1) to what degree our certifier computes useful (not too conservative) Lipschitz bounds, (2)~how much computation is required
to do so, and (3)~whether it can be usefully applied.
All reported experiments were carried out on a 2021 MacBook Pro (Model ``MacBookPro18,3'', 8 core Apple M1 Pro, 16 GB RAM, MacOS 15.2).

\subsection{Certifier Performance}\label{sec:performance}
\cref{fig:performance} depicts performance results for our certifier, measuring the usefulness of the bounds it computes and the time required
to compute them, for different numbers of Gram iterations (parameter~$n$
in \cref{fig:dafny-GramIteration}). We evaluated this against a dense ReLU MNIST~\cite{MNIST} model that comprises 8 hidden layers, each with 128 neurons.
We note that this model has a comparable number of neurons to, and significantly more layers than, the MNIST models evaluated in Kabaha et al.'s recent work on verified global
robustness properties~\cite{kabaha2024verification} (discussed later in \cref{sec:related-work}).
We trained the model using the globally robust training method of Leino et al~\cite{leino2021}, using the hyperparameters 
from the most closely related of their models~\cite[Table B.2, row 1]{leino2021} as detailed in \ifExtended\cref{app:hyper}\else{the extended~\cite{EXTENDED}}\fi. The model does not use bias terms
(because our certifier specifications do not currently handle non-zero biases; see \cref{sec:limitations}).
Leino et al.'s training method produces a model with an extra output class ``$\bot$'' that is output whenever their certification procedure
fails (i.e. decides that the model's answer was not robust).
After training the model we discard the $\bot$ output class to obtain an ordinary dense MNIST ReLU neural network, to which we apply our certifier to compute Lipschitz bounds.
That neural network we then apply to the 10,000 MNIST test points, producing 10,000 output vectors. We then
apply the certifier to those to determine the percentage certified robust at perturbation bound~$\epsilon=0.3$.
\cref{fig:performance} reports this percentage (left axis) plus the time to compute the Lipschitz bounds (right axis).

We see that increasing Gram iterations produce tighter Lipschitz bounds. At 11+ iterations, we certify
95.74\% of the 10,000 test points as robust.

To understand the quality of the Lipschitz bounds, we compare the percentage of test points that our verified certifier certifies as robust against the
percentage of test points certified robust by Leino et al.'s unverified implementation~\cite{leino2021}, which we measure directly after
training the globally robust model as the percentage of non-$\bot$ outputs when the model is applied to the 10,000 test points. We denote
this measure the model's \emph{Unverified Robustness} in \cref{fig:performance}.
This percentage (95.76\%) is just 0.02 percentage points above that of our certifier.

We also empirically compute an upper bound on the model's true robustness against the 10,000 test points by
carrying out various adversarial attacks on the model,
including FGSM~\cite{goodfellow2015explaining}, the Momentum Method~\cite{dong2018boosting}, and PGD~\cite{mkadry2017towards},
implemented using the Adversarial Robustness Toolbox~\cite{ART}. For each of the original 10,000 test points~$x$,
this gives us a set $\{x'_1,x'_2,\ldots,x'_m\}$ of perturbed points where $||x-x'_i|| \le \epsilon$.
The model's \emph{Measured Robustness} is then the proportion of
points~$x$ for which \emph{all} of the corresponding $x'_1,x'_2,\ldots,x'_m$ are classified identically to~$x$.
Our certifier's safe lower bound
on the model's robustness is 1.72 percentage points below this upper bound.

Thus our certifier produces safe robustness certifications that are extremely tight compared to
unverified (and potentially unsound) bounds.

Our certifier's performance is linear in the number of Gram iterations, because of the normalisation and truncation applied during each
Gram iteration to ensure that the sizes of the quantities involved remain roughly the same. It also benefits from the optimised, verified
implementation of the core of Gram iteration that computes~$M^T M$ (see \cref{sec:mtm}).

\subsection{Practical Usefulness}\label{sec:practical-usefulness}
%Can our certifier be usefully deployed in concert with useful models?
\emph{Verified Robust Accuracy} (VRA) measures the percentage of test points that a trained model correctly classifies and that are also
certified robust. Achieving good VRA means that
there exist useful models to which our certifier can be usefully applied. VRA also bounds a model's accuracy under
$\epsilon$-perturbations~\cite{mangal2023certifying}.
\cref{tab:performance} summarises statistics for our certifier applied to globally robust models.

We evaluated it against the
MNIST model described in \cref{sec:performance} as well as against globally robust trained Fashion MNIST~\cite{Fashion_MNIST} and CIFAR-10~\cite{CIFAR10} models
(trained with hyperparameters mimicking those of
Leino et al.~\cite[Table B.2]{leino2021}; see \ifExtended\cref{app:hyper}\else{the extended~\cite{EXTENDED}}\fi). All of these models are
dense models without bias terms and employing only ReLU activations, as required by our certifier. All are comparable in
size to, if not significantly larger than, the corresponding models considered by Kabaha et al.'s state-of-the-art work on verified global robustness properties~\cite{kabaha2024verification}.

The resulting MNIST 
model with our certifier performs very close to the unverified and potentially unsound implementation of
Leino et al.~\cite{leino2021}, with VRA within 0.01 percentage
points of their implementation applied to the same model. 
State-of-the-art (unverified) VRA for MNIST models at $\epsilon = 0.3$ is 95.7\%~\cite{leino2021} (for convolutional globally robust models employing
MinMax~\cite{anil2019sorting} activations), which is just 0.3 percentage points higher
than what we were able to achieve.

Fashion MNIST is a more challenging machine learning task than MNIST.
Our certifier can be usefully applied here
as the results in \cref{tab:performance} indicate. The 12-hidden layer, 1664-hidden neuron globally robust model we trained achieved 89.1\% accuracy, which is on par with the accuracy typically achieved by
(non-globally robust) Dense ReLU Fashion MNIST models~\cite{Fashion_MNIST_Hyperparameters}. At 12 Gram iterations our certifier was able to
compute very useful Lipschitz bounds in just 20 minutes. Its safe robustness lower bound of 83.65\% at $\epsilon = 0.25$
was within 0.05 percentage points of that computed by
Leino et al.'s unverified implementation. The resulting model and our certifier together achieved 79.54\% VRA, just 0.03 percentage points below
the unverified estimate and 6 percentage points \emph{above} the (to our knowledge) best prior VRA for Fashion MNIST models at $\epsilon = 0.25$~\cite{DBLP:conf/iclr/FromherzLFPP21}. 

CIFAR-10 is a more difficult image classification task than Fashion MNIST.
We trained a 1536-hidden neuron model, whose first two hidden layers had 512 and 256 neurons respectively, to account for this extra difficulty.
This model has twice the layers and $2.3 \times$ the neurons of the CIFAR-10 model considered by Kabaha et al.~\cite{kabaha2024verification}.
The accuracy of the
resulting model was 57.7\%, which is approx.\ 30 percentage points lower than the most advanced globally robust CIFAR-10 models~\cite{hurecipe}. Even so, the increased size of this model's inputs ($\sim 4 \times$ larger than for Fashion MNIST) means that
our certifier takes hours rather than minutes to compute tight Lipschitz bounds. At 12 Gram iterations, the
resulting VRA we obtain for $\epsilon=0.141$ is 35.95\%, which is just 0.22 percentage points below that computed by Leino et al.'s
unverified implementation. It is also $\sim$9 percentage points \emph{higher} than the (to our knowledge) best previously
reported formally
verified VRA for a CIFAR-10 model at $\epsilon=0.1$~\cite{DBLP:conf/iclr/FromherzLFPP21}. However, state-of-the-art (unverified) VRA for advanced CIFAR-10 models at $\epsilon = 0.141$ is 78.1\%~\cite{hurecipe}, which suggests that dense ReLU models may not have sufficient capacity to be  trained to be both accurate and globally robust for CIFAR-10. 

We conclude that our certifier can be practically applied to machine learning tasks for which dense ReLU robust models can be trained.

\newcommand{\conc}{+}

\begin{table}[t]
  \begin{center}
\begin{tabular}{ccccccccc}
  Dataset & \shortstack{Hidden\\Neurons} &  Accuracy & $\epsilon$  & Gram & \shortstack{Time\\(hh:mm:ss)} & \shortstack{Verified\\Robustness} & VRA \\
  \hline
% data  hidden            acc       eps   gram   time    robustness     vra
MNIST & $8 \times [128]$ & 98.4\% & 0.3 &  11 & 0:17:02 & 95.74\% (-0.02) & 95.40\%(-0.01)\\[2ex]
\shortstack{Fashion\\MNIST} &
    \shortstack{$[256] \conc$\\
     $11 \times [128]$}  & 89.1\%   & 0.25 & 12 & 0:20:08 & 83.65\% (-0.05) & 79.54\%(-0.03)\\[2ex] 
CIFAR-10 & \shortstack{%
           $[512,256] \conc$\\
       $6 \times [128]$} & 57.7\% & 0.141 & 12 & 19:02:32    & 46.12\% (-0.30) & 35.95\%(-0.22)\\
\end{tabular}

\end{center}
  \caption{\label{tab:performance}Applying the certifier. \emph{Hidden Neurons} describes the dense model architecture: $k \times [n]$ denotes $k$ hidden layers, each with $n$ neurons.
    The list $[n_1,n_2,\ldots,n_k]$ denotes~$k$ hidden layers where the $i$th hidden layer has $n_i$ neurons.
    We use $\conc$ to denote composition of hidden layers.
   $\epsilon$ is the perturbation bound at which robustness was certified over the test set.
    \emph{Gram} is the number of Gram iterations. \emph{Time}
    is the time for our certifier to compute Lipschitz bounds. \emph{VRA} is Verified Robust Accuracy. $y$\% (-$x$) denotes percentage value $y$ obtained from our certifier, which is $x$ percentage points below the unverified estimate computed by Leino et al.'s implementation~\cite{leino2021}.}
\end{table}

\section{Related and Future Work}\label{sec:related-work}\label{sec:limitations}\label{sec:future-work}

In contrast to our approach, prior work on formally verified robustness
guarantees for neural networks focuses on symbolic reasoning over the
neural network itself~\cite{katz2017,singh2019} (see e.g.~\cite{meng2022} for a survey). This has
the disadvantage that the complexity of the symbolic reasoning scales with
the size of the neural network. 

Most of this work focuses on verifying local robustness, and requiring symbolic reasoning for each point that is to be certified.
Like ours, the recent work of Kabaha et al.~\cite{kabaha2024verification} instead focuses on verifying a global robustness property.
Our work considers $l_2$ global robustness
whereas Kabaha et al. consider instead a specialised robustness property, parameterised by an input perturbation function, that considers the robustness
of a specific output class relative to the model's confidence about that class.
\iffalse % cut for space
We conjecture that our verified certifier could be applied to certify an analogous output class-based property phrased over the
verified margin Lipschitz bounds that it computes.
\fi
Kabaha et al. employ mixed-integer programming and, like other approaches that also reason symbolically over the
model, suffers similar symbolic scalability challenges~\cite{katz2017,singh2019,meng2022}. 

Our approach in contrast avoids this symbolic scalability problem entirely. It is
influenced by ideas from the field of formally verified \emph{certifying computation}~\cite{alkassar2011,rizkallah2015}: rather than trying to formally verify a complex algorithm, we instead write and formally verify a checker that certifies the outputs of that algorithm. Thus symbolic reasoning complexity no longer scales with the size of the neural network but rather with the complexity of the certification program. In our case, we base our certifier on ideas from globally-robust neural networks~\cite{leino2021}, which we augment with sound methods for computing Lipschitz bounds~\cite{delattre2023}, and all of which we formally verify in Dafny for the first time. 

Our certifier's current implementation handles a relatively simple class of neural networks,
namely dense feed-forward networks that use only the ReLU activation function.
It also does not currently handle biases, but extending it to do so would
be straightforward by extending our specification of neural network application $\ApplyNeuralNet$ (\cref{sec:neural-net-model}).
We might be able to further improve our certifier's running time to compute Lipschitz bounds by avoiding compiling Dafny's $\Real$s to arbitrary precision
rationals, instead compiling them to sound interval arithmetic~\cite{brucker2024formally}.
Even so, our certifier is still usefully applicable (\cref{sec:eval}).

Extending it to convolutional neural nets may be possible in future, leveraging ideas of~\cite{leino2021,delattre2023}.
A more interesting limitation of our approach relates to the top-level
robustness specification (\cref{sec:spec}), which
encodes neural network application with real-valued arithmetic.
\iffalse
Put another way, this specification assumes that the neural network will
be implemented using an \emph{idealised} arbitrary-precision real-valued arithmetic (e.g.\ like
that to which our robustness certifier itself is compiled).
\fi
In reality,
the neural network implementation will of course use floating point arithmetic~\cite{jin2024getting}.
\iffalse
Therefore, it is possible that the Lipschitz bounds we compute (while
valid for the idealised implementation) might be insufficiently conservative (i.e., unsound)
in practice (e.g. due to floating point rounding errors~\cite{jin2024getting}).
\fi
Closing this gap is a key avenue for future research, where we might leverage deductive verification approaches to
bounding floating point error~\cite{dross2021making}.
\iffalse
from which we can draw on
the large body of work reasoning about the gap between real and floating point
arithmetic.
\fi

\iffalse % cut for space
Finally, we note that the input parser for our Dafny program (that
takes a textual representation of the neural network and parses it to produce
its matrix representation) is currently unverified.
\fi

\section*{Acknowledgements}

This work has been supported by the joint CATCH MURI-AUSMURI.

\bibliographystyle{splncs04}
\bibliography{main}

\ifExtended % appendices only in extended version

\appendix
\section{Floating Point Unsoundness in an Unverified Certifier}\label{app:unsound-fp}

In \cref{sec:vuln-fp} we mentioned how floating point imprecision in Leino et al.'s implementation for
certifying robustness can be exploited, because it can cause their certifier to incorrectly calculate Lipschitz
constants of 0 for models with tiny weights. We avoid this issue by implementing our certifier over Dafny's
$\Real$ type, which is compiled to arbitrary precision rationals (see \cref{sec:spec}).

In this appendix, we explain the details of this vulnerability on a toy neural network for ease of
exposition. We consider a two-neuron network that takes inputs of length 1, producing output vectors of length 2.
This model thus has only two weights. We initialise the neural network with symmetric weights: \texttt{numpy.finfo(numpy.float32).tiny} and
\texttt{-numpy.finfo(numpy.float32).tiny} respectively, where \texttt{numpy.finfo(numpy.float32).tiny} $\approx 1.1754944 \times 10^{-38}$ is the smallest
positive normal \texttt{float32} value. We write \texttt{tiny} to abbreviate this value henceforth.  For an input value~$x$, this neural network
computes logits~$[\mathtt{tiny} \cdot x, -\mathtt{tiny} \cdot x]$. This neural network is designed to output
vectors whose argmax is 0 when given a positive number as its input, and whose argmax is 1 when given a negative number. It outputs the
vector $[0,0]$ when given the input~$0$.

We train this network with Leino et al.'s training algorithm for 100 epochs, using training data that ensures the initial model weights
remain unchanged. Specifically, we train under sparse categorical cross-entropy loss using
two training examples: $\texttt{tiny} \mapsto 0$ and $-\texttt{tiny} \mapsto 1$ (where we write $x \mapsto y$ to denote a training input~$x$
whose true label is~$y$). Because both the training inputs and weights are tiny, the model computes logits for both training samples of $[0,0]$. As a result,
the derivatives of the loss wrt the logits are symmetric and $< 0$. Because the weights mirror this symmetry and are tiny, the derivatives of the loss wrt the weights
end up being calculated as $0$. Thus training does not update the model weights and they stay tiny.

During training,
Leino et al.'s implementation estimates the model's Lipschitz bounds from the (unchanged) 
tiny weights.
Due to floating point imprecision,
Leino et al.'s implementation produces margin Lipschitz bounds of 0, which are obviously incorrect. Their implementation then incorrectly certifies \emph{all} outputs as robust against \emph{all} perturbation bounds! In fact, this neural network is highly non-robust (e.g., consider the region around the output point $[0,0]$, for input $0$). Therefore, this example invalidates the soundness of Leino et al.'s certified robustness implementation.

In contrast, our verified certifier correctly refuses to certify the output $[0,0]$ even for the trivial perturbation bound of $\epsilon = 0$. 
The margin Lipschitz constants for this neural network are each bounded above by the $l_2$ norm of the single-element vector that is the subtraction of
the two weights (see \cref{sec:cauchy-schwartz}). This means they are bounded by \mbox{$2 \cdot \texttt{tiny} \approx 2.3509887 \times 10^{-38}$}. Our certifier computes safe margin
Lipschitz bounds of $7.2761\times10^{-12}$ for this neural network.

\section{Normalisation Unsoundness in an Unverified Certifier}\label{app:unsound-normalization}

In \cref{sec:vuln-normalisation} we mentioned an exploitable implementation flaw (bug) in Leino et al.'s certifier
implementation that causes it to under-estimate Lipschitz constants for models with small weights.
This issue
occurs due to a subtly unsound implementation of normalization as part of the power method (see \cref{sec:power}).

As mentioned in \cref{sec:power},  the power method involves iterating on a vector $\vv_i$ by applying the recurrence
$\vv_{i+1}=M^T M\vv_i$. After each iteration, normalisation of the resulting vector is performed to prevent overflow.
For vector~$\vv$ its $l_2$ normalisation is $\frac{1}{||\vv||} \cdot \vv$. When the $l_2$ norm is zero, however, one must take care to
avoid division by zero. Leino et al.'s implementation does so by adding a small quantity to~$||\vv||$. We denote this small
quantity~$e$. Thus to normalise a vector~$\vv$, their implementation computes $\frac{1}{||\vv||+e} \cdot \vv$. The value of~$e$ in Leino et al.'s implementation is  $e=1\times10^{-9}$; however the issue we describe here does not depend on the
specific value of this quantity.

Unfortunately, when $||\vv_{i+1}||$ is non-zero but significantly smaller than~$e$, the addition of this quantity means that repeated normalization
has the effect of artificially reducing $||\vv_{i+1}||$. Thus Leino et al.'s implementation can end up significantly under-estimating the Lipschitz
constants of non-final layers of the model (to which power iteration is applied).

For instance, consider a two-layer neural network. As in \cref{app:unsound-fp}, this neural network takes single-element vectors as its
input and outputs 2-element vectors. Its first layer has a single neuron that is initialised with the weight equal to $1 \times 10^{-5}$.
Let~$w = 10^{-5}$ denote the first layer's sole weight.
Its second layer
has just two neurons with weights $[1.0, -1.0]$. Training this neural network on the same training data as in \cref{app:unsound-fp} does
not cause its weights to change. Therefore, the Lipschitz constant for the first layer is simply~$w$.

When performing power iteration for the first layer of this neural network, $M = [w]$. Thus $\vv_{i+1}=w^2 \cdot \vv_i$. Assuming $\vv_i$ is
properly
normalised (i.e.,\ its length is 1), we see that $||\vv_{i+1}|| = w^2 = 1 \times 10^{-10}$ which is significantly less than~$e = 1 \times 10^{-9}$.

As a result,
Leino et al.'s implementation of the power method estimates the Lipschitz constants for the first layer as $9.434563 \times 10^{-6}$ which is
below~\mbox{$w=10^{-5}$}, i.e., below the actual Lipschitz constant for this layer. It then estimates the neural network's margin Lipschitz bounds unsafely, as
$1.8869127 \times 10^{-5}$, when this value should be at least $2w = 2 \times 10^{-5}$ to be safe.
Leino et al.'s implementation then incorrectly certifies non-robust outputs. 
In contrast, our verified certifier calculates safe Lipschitz bounds for this neural network of $\sim 2.002w$ and produces sound certifications.

Together with unsoundness due to floating-point precision (\cref{app:unsound-fp}), this implementation soundness issue highlights
the value of formal verification for robustness certification. We conclude that Leino et al.'s implementation is sufficient for
\emph{training} globally-robust neural networks; however, a verified certifier should be applied to then certify the outputs of such networks
when deployed in safety- and security-critical applications.

\section{Certification Unsoundness in an Unverified Certifier}\label{app:unsound-neginf}

In \cref{sec:vuln-certification} we mentioned an exploitable vulnerability in
Leino et al.'s implementation for certifying the robustness of individual output points. Here we describe that issue
in detail, with reference to a toy neural network to ease exposition.

Leino et al.'s certifier implementation works like a wrapper that augments a neural network with an
additional output logit. We say that their implementation \emph{wraps} the underlying neural network.
This additional logit is denoted $\bot$.  Its value is intended to be computed such that whenever the wrapped model's output is not robust, the $\bot$ logit's value dominates all other logits.

Suppose the wrapped neural network produces outputs~$y$ of length~$n$. Then Leino et al.'s implementation produces
output vectors of length~$n+1$, with the additional~$\bot$ logit, whose value we denote~$y_\bot$.
Having obtained the output vector~$y$ from the wrapped neural network, $y_\bot$ is computed as follows.
Letting $j$ denote $\ArgMax(y)$ and $y_j$ denote $y[\ArgMax(y)]$,
first a vector~$z$ of length~$n$ is computed such that each element~$z_i$ is equal to $y_i$ + $\epsilon \cdot L_{i,j}$. The \emph{intention} of Leino et al.'s implementation is then to replace the $j$th entry of~$z$
with negative infinity. $y_\bot$ is then the maximal value of the resulting vector. What their implementation does instead is to compute a vector~$m$
of length~$n$ such that each element~$m_i$ of~$m$ is equal to $-\mathtt{inf}$ whenever $y_i$
is equal to $y_j$, and is $z_i$ otherwise. ($y_\bot$ is computed as the maximal element of~$m$.)
This replaces not only~$z_j$ with negative infinity but also any other elements~$z_i$ whose output logit~$y_i$
happens to be equal to~$y_j$.

Unfortunately, this leads to incorrect computation of~$y_\bot$ and
unsound certification results for neural networks that output vectors~$y$ with equal
logits. 

Consider the same neural network as in Appendix~\ref{app:unsound-normalization} except the sole weight of
its first layer is 0.9 (instead of $1 \times 10^{-5}$). Then the decision boundary for this neural network remains at 0: it
outputs vectors whose argmax is 0 for all non-negative inputs, and 1 otherwise. For the input~$0$, it produces the
output vector $y = [0,0]$ whose argmax~$j = 0$ (breaking ties by taking the first index, as in all standard implementations) and
of course $y_i = y_j$ for all~$i$.
Thus we have that all $m_i$ are $-\mathtt{inf}$ and so $y_\bot$ is incorrectly computed as negative infinity (when instead it
should be $0 + \epsilon \cdot L_{i,j} = 1.8\epsilon$). Thus the output $[0,0]$ is incorrectly certified robust for all $\epsilon$ even though
it is produced by the input~$0$ which is precisely this model's decision boundary (and so non-robust by definition).

This issue further demonstrates the need for a formally verified certification routine, like ours.

\section{Matrix Operations in Dafny}\label{sec:appendix-matrix-ops}
%\section{Appendix}

In this appendix, we outline the formal specifications of several basic matrix operations, which we specify over matrices and vectors represented as Dafny sequences. This allows us to specify these operations as pure functions. In our certifier's implementation,
many of these specifications are implemented imperatively. Key operations like the matrix-matrix product~$M^T M$ of the transpose~$M^T$ of
a matrix~$M$ with itself are implemented by specialised routines that operate over two-dimensional arrays of $\Real$s, rather
than Dafny sequences, to maximise efficiency. Naturally those implementations are verified correct against their functional
correctness specifications that follow.

%\subsection{Matrix Operations}\label{sec:appendix-matrix-ops}

\begin{flalign*}
\TotalSpec
  {\FUNCTION
    {\Sum}
    {v:\Seq{\Real}}
    {\Real}
  }
  {\ITE{|s|=0}{0\ }{\Sum(s[..|s|-1])+s[|s|-1]}}
\end{flalign*}
\begin{flalign*}
\TotalSpec
  {\FUNCTION
    {\Square}
    {x:\Real}
    {\Real}
  }
  {x\cdot x}
\end{flalign*}

\begin{flalign*}
\TotalSpec
  {\FUNCTION
    {\GetFirstColumn}
    {M:\Matrix}
    {\Vector}
  }
  {\\&\quad\ITE{|M|=1}{[M[0][0]]\ }{[M[0][0]]+\GetFirstColumn(M[1..])}}
\end{flalign*}
\begin{flalign*}
\Spec
  {\Cols(M)>1}
  {\FUNCTION
    {\RemoveFirstColumn}
    {M:\Matrix}
    {\Matrix}
  }
  {\\&\quad\ITE{|M|=1}{[M[0][1..]]\ }{[M[0][1..]]+\RemoveFirstColumn(M[1..])}}
\end{flalign*}
\begin{flalign*}
\TotalSpec
  {\FUNCTION
    {\Transpose}
    {M:\Matrix}
    {\Matrix}
  }
  {\\&\quad\ITE{\Cols(M)=1}{[\GetFirstColumn(M)]\\&\quad}{[\GetFirstColumn(M)]+\Transpose(\RemoveFirstColumn(M))}}
\end{flalign*}

\begin{flalign*}
\TotalSpec
  {\FUNCTION
    {\FrobeniusNorm}
    {M:\Matrix}
    {\Real}
  }
  {\\&\quad\Sqrt(\SumMatrixElements(\SquareMatrixElements(M)))}
\end{flalign*}
\begin{flalign*}
\TotalSpec
  {\FUNCTION
    {\SumMatrixElements}
    {M:\Matrix}
    {\Real}
  }
  {\\&\quad\ITE{|M|=1}{\Sum(M[0])\ }{\Sum(M[0])+\SumMatrixElements(M[1..])}}
\end{flalign*}
\begin{flalign*}
\TotalSpec
  {\FUNCTION
    {\SquareMatrixElements}
    {M:\Matrix}
    {\Real}
  }
  {\\&\quad\ITE{|M|=1}{[\Apply(M[0],\Square)]\\&\quad}{[\Apply(M[0],\Square)]+\SquareMatrixElements(M[1..])}}
\end{flalign*}

\begin{flalign*}
\Spec
  {\Cols(M)=|v|}
  {\FUNCTION
    {\MV}
    {M:\Matrix,\ v:\Vector}
    {\Vector}
  }
  {\\&\quad\ITE{|M|=1}{[\DotProduct(M[0],v)]\\&\quad}
  {[\DotProduct(M[0],v)]+\MV(M[1..],v)}}
\end{flalign*}
\begin{flalign*}
\Spec
  {|v|=|u|}
  {\FUNCTION
    {\DotProduct}
    {v:\Vector,\ u:\Vector}
    {\Real}
  }
  {\\&\quad\ITE{|v|=1}{v[0]\cdot u[0]\\&\quad}
  {v[0]\cdot u[0]+\DotProduct(v[1..],u[1..])}}
\end{flalign*}

\begin{flalign*}
\Spec
  {\Cols(M)=\Rows(N)}
  {\FUNCTION
    {\MatrixMatrixProduct}
    {M:\Matrix,\ N:\Matrix}
    {\Matrix}
  }
  {\\&\quad\ITE{|M|=1}{[\MMGetRow(M[0],N)]\\&\quad}{[\MMGetRow(M[0],N)]+\MatrixMatrixProduct(M[1..],N)}}
\end{flalign*}
\begin{flalign*}
\Spec
  {|v|=|N|}
  {\FUNCTION
    {\MMGetRow}
    {v:\Vector,\ N:\Matrix}
    {\Vector}
  }
  {\\&\quad\ITE{\Cols(N)=1}{[\DotProduct(v,\GetFirstColumn(N))]\\&\quad}{[\DotProduct(v,\GetFirstColumn(N))]\\&\quad+\MMGetRow(v,\RemoveFirstColumn(N))}}
\end{flalign*}

\section{Model Training Hyperparameters}\label{app:hyper}

The table below details the hyperparameters used to train the models in \cref{sec:eval} using Leino et al.'s globally robust neural networks
training algorithm~\cite{leino2021}. We refer to Leino et al.~\cite{leino2021} and their training implementation\footnote{\url{https://github.com/klasleino/gloro/tree/master/tools/training}}  for the meaning of each hyperparameter and value, while noting that
the hyperparameter choices were made to mimic those used in Leino et al.'s evaluations~\cite[Table B.2]{leino2021} as closely as possible.
``CE'' stands for sparse categorical cross-entropy loss.
Since Leino et al. did not evaluate a Fashion MNIST model, we base its hyperparameter choices on those for our MNIST model, increasing the
batch size to keep acceptable model training time, and decreasing $\epsilon_{\mathit{train}}$ to account for the increased difficulty of
this learning task over MNIST.

\bigskip

\noindent\begin{tabular}{ccccccccccc} 
  model & \shortstack{\#\\epochs} & \shortstack{batch\\size} & loss & $\epsilon_{\mathit{train}}$ & \shortstack{\emph{initial-}\\\emph{ization}} & \emph{init\_lr} & \emph{lr\_decay} & \emph{$\epsilon$\_schedule} & \shortstack{\emph{augment-}\\\emph{ation}} \\[1ex]
  \hline\\[1ex]
  MNIST & 500 & 32 & CE & 0.45 & default & 1e-3 & \shortstack{decay\_to-\\\_1e-6} & single & none \\[2ex]
  CIFAR-10 & 800 & 256 & CE & 0.1551 & default & 1e-3 & \shortstack{decay\_to-\\\_1e-6} & single & none \\[2ex]
  \shortstack{Fashion\\MNIST} & 500 & 64 & CE & 0.26 & default & 1e-3 & \shortstack{decay\_to-\\\_1e-6} & single & all
\end{tabular}

\fi % ifExtended for appendices

\end{document}